%Paper: hep-th/9402028
%From: Stam Nicolis <nicolis@marcptsu3.univ-mrs.fr>
%Date: Fri, 4 Feb 94 14:50:34 +0100

\magnification=1200
\catcode `@=11

\hsize 17truecm
\vsize 24truecm
\font\twelve=cmbx10 at 13pt
\font\ten=cmbx12
\font\eightrm=cmr8

\baselineskip 15pt
%\nopagenumbers

\centerline{\ten CENTRE DE PHYSIQUE THEORIQUE}
\centerline{\ten CNRS - Luminy, Case 907}
\centerline{\ten 13288 Marseille Cedex 9 - France}
\centerline{\bf UPR 7061}

\vskip 4truecm

\centerline{\twelve Z(N) BUBBLES AND THEIR DESCENDANTS}
\centerline{{\twelve  IN HOT QCD~: ALIVE
AND BUBBLING\footnote{$^{\displaystyle\star}$}{\eightrm Talk
given at the Banff Conference on Thermal Field Theory, August
12-25, 1993.}}}

\bigskip

\centerline{\bf C.P. KORTHALS ALTES}

\vskip 3truecm

\centerline{\bf Abstract} 

\medskip

The spontaneous breaking of Z(N) symmetry in hot QCD
and the appearance of domain walls is reviewed.

\vskip 4truecm

%\noindent Number of figures : 5

\bigskip

\noindent January 1994

\noindent CPT-94/P.3006

\bigskip

\noindent anonymous ftp or gopher : cpt.univ-mrs.fr

\vfill\eject

%\end

%1st updating with effect from: 5 Sept 1991

%2ND UPDATING WITH EFFECT FROM: 28 JUNE 1993
%(for the purpose of making PlainTex file + Latex file identical)

%--------------------------------------------------------------------
\headline={\ifnum\pageno=1\firstheadline\else
\ifodd\pageno\rightheadline \else\leftheadline\fi\fi}
\def\firstheadline{\hfil}
\def\rightheadline{\hfil}
\def\leftheadline{\hfil}
%	\footline={\ifnum\pageno=1\firstfootline\else\otherfootline\fi}
%\def\firstfootline{\rm\hss\folio\hss}
%\def\otherfootline{\hfil}

\font\twelvebf=cmbx10 scaled\magstep 1
\font\twelverm=cmr10 scaled\magstep 1 
\font\twelveit=cmti10 scaled\magstep 1

\font\tenrm=cmr10

\font\eightrm=cmr8

\parindent=1.5pc
%\hsize=6.0truein
%\vsize=8.5truein
%\nopagenumbers

\hoffset=0truecm
\voffset=1truecm

\vfil
\twelverm
\baselineskip=14pt
\leftline{\twelvebf 1. Introduction}
\vglue 0.4cm

In recent years much attention has been paid to the quark-gluon
plasma, both numerically$^1$ and analytically$^2$. The advent of RHIC
at Brookhaven and  possibly heavy ion experiments at LHC and the
relevance for early universe  scenario's amply justify its study.

The purpose of these notes is to provide some additional insights
into the  phenomenon of spontaneous Z(N) breaking$^3$ at high
temperatures, and to  provide an up to date presentation of what has
been achieved. I would like to apologise right from the start to the
lattice people~: they always asked the relevant questions and are not
really concerned with the partly rather academic questions posed in
this talk.

Z(N) is the centergroup of the SU(N) gauge group and as such does not
appear in the gauge transformation law of the gauge potentials. The
symmetry appears only at the quantum level~: physical states do
transform according to irreducible representations of the
centergroup. When we compute the free energy of the system we find a
low temperature phase where the free energy is non-degenerate. In the
high temperature phase there  appears a degeneracy.        

Traditionally the phenomenon of broken Z(N) symmetry is described
in close analogy with spontaneous magnetisation in spin systems$^3$;
in particular one expects the occurence of ordered phases, and domain
walls in between them. Already at this point the  reader may raise
his eyebrows~: how can there be a close analogy between  spontaneous
magnetisation, a typical low temperature phenomenon, and the  would
be spontaneous breaking of Z(N) symmetry at {\twelveit high}
temperatures~? In other words,  how can it possibly be, that a
symmetry breaks at high temperatures~? The answer is that the latter
is perfectly possible, once we realise that we  carry a prejudice on
how the surface tension $\alpha$ of a small bubble of  say down
spins behaves in a large sea of up spins. We usually think of it as
being temperature independent at low temperatures. That means that
the probability of having such a droplet of radius R at temperature T
equals~:
$$exp{-{\alpha\over T} R^2}$$
  
Given our prejudice on $\alpha$ this probability disappears at small
T, and  an ordered phase prevails. However in QCD without quarks the
only scale is  T itself~! Hence dimensional arguments tell us that
$\alpha\sim T^3$ and now the scenario is reversed~: at {\twelveit
high} temperatures order installs itself~!\footnote*{\tenrm I learnt
this argument from R.D.~Pisarski}. Actually the surface tension has
been computed in perturbation theory$^2$ with results that were in
qualitative agreement with Montecarlo results$^1$.

There remains another, somewhat mystical element in the argument
above	: what are the degrees of freedom in QCD that do correspond to
these spins~? They are the expectation values of the Polyakov lines
$P(A_0)$ and are related$^3$ to the excess free energy of a single
heavy test quark~:
$${Tr\langle q\vert\exp{-{H\over T}}\vert q\rangle\over
{Tr\exp{-{H\over T}}}}=\langle P(A_0)\rangle$$
In this formula the trace is over physical states, and q represents
the heavy quark. The average on the right hand side is the Euclidean
path integral average. The point is now~: make a gauge transformation
periodic up to a Z(N) phase in the time direction, and the Polyakov
loop will pick up this phase (of course the action in the path
integral stays invariant). On the other hand the free energy of a
quark cannot acquire such a phase, and the way out is that both
averages are zero. This does {\twelveit  not} imply, that Z(N)
symmetry never breaks. It just reflects the basic fact, that the
order parameter can {\twelveit only} become non-zero  once we
trigger it by appropriate boundary conditions, as is well known in
spin systems. Incidentally, the fact that the Hamiltonian matrix
element above is zero, can be deduced from the fact that the one
quark state can never be made gauge invariant, since Gauss' law cannot
be satisfied.

Of course, lattice people have been undaunted by this. They measured
two point functions, and showed, that above a critical temperature the
confining behaviour was replaced by a screened Coulomb
potential$^3$. But to show the presence of a surface tension between
coexisting phases, we have to be careful and specify boundary
conditions. In section 3 we will do so. It will become clear, that
the relative phase of Polyakov loops is perfectly physical. To
illustrate the reasoning we have inserted a section on the paradigm~:
${\varphi}^4$ theory in three dimensions. The analytic computations in
QCD are discussed in section 4. Section 5 is a summary and an outlook.

\vskip 0.6cm
\leftline{\twelvebf 2. A useful finger exercise~: ${\varphi}^4$
theory} 
\vglue 0.4cm

The prime field theory example of a spin system with a Z(2) symmetry
is as we said the one with the three dimensional action~:
$$S=\int d\vec x \big({1\over 2}\sum_k (\partial_k\varphi)^2 +
{\lambda\over 4!}({\varphi}^2 -v^2)^2\big)\eqno 2.1$$
The symmetry of the action is $\varphi \to -\varphi$. We have written
the potential such that the $Z(2)$ symmetry is spontaneously broken
(see fig. 1), the vacuum expectation value of the field being  $\pm
v$. The latter is related to the mass term ${-m^2\over 2}{\varphi}^2$
by $v=\sqrt{6{m^2\over\lambda}}$.

The physics is identical in either of the two minima. One can think of
a box with sizes $L\times L\times L_z$, 
with $L_z>>L>>m^{-1}$, in which we trigger
the two phases by suitable boundary conditions (see fig. 2). The
region  in between the two coexisting phases is having a
higher energy density due to the appearance of a domain
wall
$$\varphi_c(z)\sim tghmz \eqno 2.2$$
as follows from minimising
the action eq.~2.1 under the condition that
$\varphi=\pm v$ at $z=\pm\infty$. The scale over which this wall
develops is $m^{-1}$. The actual total energy stored in the wall
per unit of surface is
$$\alpha_0=4\sqrt 2 {m^3\over {\lambda}}\eqno 2.3$$
How can we see the wall, other than by its energy density~? The obvious answer
is~:
take the two point function $\langle\varphi(x) \varphi(y)\rangle$,
with y fixed in say the plus phase, and x varying from the plus to the
minus phase. In this trivial example there are no fluctuations so the
correlation will
trace out the profile eq.~2.2, and becomes for large enough
separation  negative (see fig.~2).

\vskip 0.4 cm
\leftline{\twelveit 2.1 Quantum corrections}
\vglue 0.3 cm

To compute the quantum corrections we have to resort to the path
integral. We will take boundary conditions to be periodic in x and 
y directions and (anti)periodic in the z direction and the
corresponding
partition functions will be called 
$(Z_{(0,0,1)})\break Z_{(0,0,0)}$.
To make sense of the path integrals we will put the action eq.~2.1 on
the lattice, by discretising the derivatives in the kinetic term~:
$$\partial_k \varphi(\vec x)\equiv {\varphi(\vec x_+\vec i_k
a)-\varphi(\vec x)\over a}$$
a being the lattice length.
This means that the kinetic term becomes
$$-a^{-2}\sum_k\varphi(\vec x +\vec i_ka)\varphi(\vec x)\eqno2.4$$ 
apart from local terms. The anti-periodicity in the z direction can be
reformulated simply by absorbing the - sign at z=$L_z$ into the
{\twelveit coupling} in eq.~2.4, and to consider the field to be
periodic. The new action with this 2 d surface of anti ferromagnetic
couplings is called
``twisted".

The ratio of the twisted partition funtion over the untwisted one
will cancel out the volume effects and leaves us with  the effect of
the twist~:  
$${Z_{(0,0,1)}\over Z_{(0,0,0)}}=E(L)\exp{-\alpha L^2}\eqno 2.5$$
Here $\alpha$ is the surface energy (the prefactor E contains powers
of L). Classically it reduces to eq.~2.3, as will become explicit in
the next section. It is amusing to note that eq.~2.5 can be seen as
describing a disorder phenomenon, with a rapid fall-off. On the other
hand the correlation function will signal an ordered behaviour~:
surface tension and correlation are dual aspects of the system. In
fig. 3 we show how the correlation in the presence
of the twist picks up a - sign when piercing the twisted links. This
is seen by redefining the field variable with a minus sign where the
correlation crosses, a manifestation of the algebra proposed in
ref.~5.

\vskip 0.4 cm 
\leftline{\twelveit 2.2 Calculation of the surface energy}
\vglue 0.3 cm

Let us introduce the constrained effective action $U([\phi (z)])$~:
$$\exp{-L^2U([\phi (z)])}\equiv
\int D\varphi\prod_z\delta(\phi(z)-\bar\varphi(z))\exp{-S(\varphi)}
\eqno 2.6$$
We wrote $\bar\varphi(z)\equiv {1\over {L^2}}\int dx dy
\varphi (x,y,z)$, so up to normalisation the effective potential U
gives us the probability that a configuration with profile $\phi (z)$
appears. S is the untwisted action. The ratio in eq.~2.5 is related
very simply to the effective potential~:
$${Z_{(0,0,1)}\over {Z_{(0,0,0)}}}= \int D\phi(z)\exp{-L^2U([\phi
(z)])}\eqno 2.7$$
The only way that the anti periodicity enters in eq.~2.7 is through
the profile~: $\phi (0)=-\phi (L_z)$. 

{}From eq.~2.5 and 2.7 it follows that knowledge of the effective
action is enough to compute the surface energy. The former can be
computed from eq.~2.6 in a loop expansion~: 
$$U([\phi])= S([\phi])+ U_1([\phi])+ higher\ loops\eqno 2.8$$
S is the classical action, $U_1$ is the one loop contribution.

For large L we only have to minimise U, as follows from eq.~2.7. 
The classical term has been minimised in the previous sub-section.
The one loop contribution is computed$^4$ from the propagator with
inverse $S^{\prime\prime}(\varphi_c)$. This inverse has the correct
spectrum (i.e. no negative eigenvalues). When letting the lattice
length go to zero, we have to reexpress the parameters in the action
in terms of their renormalised counterparts.  

At small coupling, ${\lambda\over m}<<1$, the vacuum expectation
value v provides a much smaller scale, over which the wall, eq.~2.2,
develops adiabatically, like $\sqrt{\lambda} vz$.    
Nether\-theless, we are not allowed to make a derivative expansion,
since $S^{\prime\prime}(\phi)$ for a {\twelveit constant}
profile $\phi$ develops negative eigenvalues, as the potential in
fig.~1 shows, when it is convex.

This is the prime distinction with the interface in hot gauge theory
(see section~4).

\vskip 0.6cm
\leftline{\twelvebf 3. The free energy in QCD, string tension and
surface tension}
\vglue 0.4cm

In this section we will give a definition of the surface tension
using the Hamiltonian approach. The use of Euclidean path integrals
will be completely avoided, until the actual evaluation. The method
we use is  essentially an adaptation of 't Hooft's method$^5$ and a
short version was published some time ago$^6$ by us.

\vskip 0.4 cm
\leftline{\twelveit 3.1 Surface tension from the Hamiltonian}
\vglue 0.3 cm

The Hamiltonian is in a familiar notation~:
$$H=\int d\vec x Tr(g^2{\vec E}^2+{1\over{g^2}}{\vec B}^2) $$
with g the bare coupling. The electric and magnetic field strengths are written
in NxN matrix form~:
$$E_k =E^a _k\lambda^a$$
and 
$$Tr\lambda^a\lambda^b={1\over 2}\delta^{a,b}$$

The indices a, b are the $N^2-1$ colour indices. As before we
imagine the system in a box of macroscopic size $L_xL_yL_z$, with 
the size $L_z$ in the z direction much larger than the tranverse
ones. The gauge potentials have  periodic boundary conditions. The
system is put in a heatbath at temperature T.

We will now introduce symmetry operators $\Omega_{\vec k}$, commuting
with $H$. They are gauge transformations in the box which are
periodic up to a centergroup element $exp{ik_x{2\pi\over N}}$ in the x
direction and analogous for y and z  direction. Thus they will leave
gauge potentials periodic, since the Z(N)  phase does not appear in
the gauge transform of $A_{\mu}\equiv  A_{\mu}^a\lambda^a$~:

$$A_{\mu}^{\Omega}=\Omega A_{\mu}\Omega^{-1}+i\Omega\partial_{\mu}
\Omega^{-1}$$
If we define physical states as states obtained by
averaging over all truly perio\-dic gauge transformations, then these
physical states fall into $N^3$ different Hilbert spaces, labeled by
electric flux $\vec e$.  Each space is invariant under
the Hamiltonian, and any state vector $\vert\vec e \rangle$ is
eigenvector of $\Omega_{\vec k}$ with eigenvalue $exp{i\vec k.\vec e
{2\pi\over N}}$. It is then obvious that we can obtain any electric
flux state from the no flux state $\vert \vec e =\vec 0 \rangle$
by acting on it with the Wilson line\footnote*{\tenrm In our
terminology a Wilson line is a Polyakov line, eq.~3.6, in one of the
space directions} winding the appropriate number of times in the
spatial directions, because a  Wilson line $U_x$ in the x- direction
will pick up a phase $exp{ik_x{2\pi\over N}}$~:
$$\Omega_{\vec k}U_x\Omega_{\vec k}^{-1}=\exp{ik_x {2\pi\over N}}U_x$$

We are now in a position to write down the free energy of a given
electric flux~$\vec e$~:
$$\exp{-{F_{\vec e}\over T}}\equiv Tr_{phys}\langle\vec
e\vert\exp-{H\over T}\vert\vec e\rangle \eqno 3.1$$
  
The trace in eq.~3.1 is over all physical states with electric flux
$\vec e$ and it is not hard to see that any such state with a given
elctric flux can be written as 
$$|\vec e>=P_{\vec e}|phys>$$
where the projector is related to the gauge
transformations $\Omega_{\vec k}$ by
$$P_{\vec e}={1\over N^3}\sum_{\vec k}\exp{i\vec k.\vec e {2\pi\over
N}} \Omega _{\vec k}\eqno 3.2$$
For large spatial volume  volume and low enough temperature we find
for the  diffe\-rence (we want to get rid of volume effects, and only
study effects due to  the Wilson line)~:
$$F_{(0,0,1)}-F_{(0,0,0)}=\rho(T)L_z\eqno 3.3$$
whereas for high enough temperatures we will find in perturbation
theory~:
$$F_{(0,0,1)}-F_{(0,0,0)}=D(T,L)\exp{-{\alpha(T)\over T}L^2}\eqno
3.4$$ The prefactor D consists of powers of L and T.

Of course the low temperature relation is telling us that the string
tension $\rho$ will cause a linearly rising energy difference between
the various free energies and that $F_{(0,0,0)}$ is the lowest energy
state.

The high temperature relation renders the free energies all degenerate
in the the infinite volume limit, and the quantity $\alpha$ that
controls the exponential decay of the difference is the surface
tension, as we will see at the end of this section. We have here a
definition of the surface tension without introducing path integrals
or even Polyakov loops. We will only use those when  computing 
$\alpha$ in perturbation theory

\vskip 0.4 cm
\leftline{\twelveit 3.2 From Hamiltonian to path integral}
\vglue 0.3 cm

The transition from the Hamiltonian expression for the free energy,
eq.~3.1, to a path integral expression is provided by eq.~3.2. Indeed,
after  substution of eq.~3.2 into eq.~3.1, we are left with a linear
combination of  terms of the type~:
$$<phys|\exp{-{H\over T}}\Omega_{\vec k}|phys>\eqno 3.5$$
 
Were it not for the presence of the gauge transformation
$\Omega_{\vec k}$ this matrix element would equal the four dimensional
path integral with periodic boundary conditions in all directions,
including the Euclidean time direction $\tau$, that runs from $0$ to
${1\over T}$. The presence of $\Omega_{\vec k}$ in eq.~3.5 renders
the periodicity twisted. To illustrate this we take a specific $\vec k
=(0,0,1)$. At any point with $\tau=0,z=0$ we follow first say the
z-direction, and pick  up the non-periodic gauge transformation at
$\tau=0,z=L_z$. Then we go from time $\tau=0$ to $\tau={1\over T}$ and
pick up a periodic transform. Thus we find in the
corner $\tau={1\over T},z=L_z$ a gauge potential, which is  up to a
non periodic gauge transform the same as we stated from in the
opposite corner. Following the path first along the time and then
along the z-direction leaves us with a  periodic
gauge transformation. Either path leaves us with one and the same
gauge potential in the corner, because it does not feel the
centergroup.
          
What is the physical meaning of such a twisted matrix element~? First
we note that the Polyakov loops $P(\vec x)$ defined by
$$P(\vec x)\equiv {1\over N}TrP\exp{i\int_0^{{1\over T}}d\tau A_0(\vec x,
\tau)} \eqno 3.6$$
are periodic modulo $exp{i{2\pi\over N}}$ in the z-direction.
Thus the Polyakov loops have the same periodicity due to the twist,
as the spin variables in the previous section, and the twisted matrix
element has to be considered as the analogue of the anti-periodic
partition function in the spin case. Its lattice version has twisted
plaquettes (instead of links as in section 2.1.)

Let us study the ratio 
$$Z_{(0,0,1)}/Z_{(0,0,0)} \eqno 3.7$$
As in eq.~3.3, this ratio contains no volume effects.

In the low temperature phase it is related to the string tension in
eq.~3.3 by strong coupling methods$^7$. In the high temperature phase
one can use continuum perturbation theory methods to evaluate the
ratio and the outcome is
$${Z_{(0,0,1)}\over Z_{(0,0,0)}}=\exp{-{\alpha\over T}L^2}\eqno 3.8$$

Like in section 2 for the spin system, the surface tension is dual to
the correlation of two Polyakov loops, and an algebra like in ref. 5
can be established.

We have computed the surface tension to two loop order, as will be
explained in the next subsection. Accepting eq.~3.8 for the moment,
it is not hard to show from eq.~3.2, 3.4 and 3.8, that 
$$F_{(0,0,1)}-F_{(0,0,0)}=2T(1-cos{2\pi\over N}) {Z_{(0,0,1)}\over 
Z_{(0,0,0)}}\eqno 3.9$$
up to exponentially smaller terms.

\vskip 0.4 cm
\leftline{\twelveit 3.3 Computing the surface tension} 
\vglue 0.3 cm

The ratio of the two path integrals can be written as 
a path integral over all possible Polyakov-loops p(z), with as
integrand the constrained effective potential $U([p(z)]) $.
$${Z_{(0,0,1)}\over Z_{(0,0,0)}}\equiv \int Dp(z)\exp{-L^2{U([p])\over
T}}\eqno 3.10$$
The latter is defined as
$$\exp{-L^2{U([p])\over T}}\equiv \int DA_0 D\vec A\delta(p-\bar p(A_0))
\exp{-{1\over g^2}S(A)}\eqno 3.11$$
We dropped for notational convenience the z-dependence in the delta
function cons\-traint on the averaged Polyakov loop~:
$$\bar p(A_0(z))\equiv {1\over L^2}\int dx dy P(A_0(\vec x)) \eqno
3.12$$
The boundary conditions on the loop in eq.~3.10 are of course anti
periodic as  mentioned in the discussion of the twisted matrix
element. The analogy with the spin case in section 2.2 is obvious.
The constrained effective potential is independent of the boundary 
conditions, and is readily accessible in MonteCarlo
simulations$^{8,1}$.

If we are not interested in finite size effects due to L, the ratio
in  eq.~3.10 is determined by the saddle point of U([p]). The value
of U in the saddle point $\tilde p$ is the surface tension $\alpha$,
according to eq.~3.8.

So we are left with the task of calculating U from eq.~3.11. This is
done in section 4.
% modified by CLee 23/07/93

\vskip 0.6cm
\leftline{\twelvebf 4. Results of the perturbative calculation}
\vglue 0.4cm

We have calculated the effective potential eq.~3.11 up to two loop 
order$^{2,9}$. The method is well known; determine the saddlepoint of
the integrand in eq.~3.11, and expand around it, using a background
gauge fixing. The important difference with $\varphi^4$ theory is that
we {\twelveit have} to use the gradient expansion method~!

The reason for this is the following. In the classical action there
is no scale. There is only a gradient term, {\twelveit no} potential
term like in ${\varphi}^4$ theory. The way the only scale in the 
theory, the temperature, comes into the effective potential is through
the 1-loop result$^{10}$ $V_1$, which is order $g^2$ smaller than the
gradient term.  To make both of the same order, we have to admit
{\twelveit gradients of order} g. 

The effective potential, to this order, takes the form~:
$$U(p)={1\over g^2}\int dz(1+g^2K_1(q))({\partial\over \partial
z}q)^2 +\int dz(V_1(q)+g^2V_2(q))\eqno 4.1$$
The variable q is defined by
$$A_0=diag ({q\over N}, {q\over N},....,-{(N-1)\over N}q)$$
In fig. 4 we have depicted the 1-loop effective potential $V_1$ for
the case N=3, with global minima at the Z(3) values of
the Polyakov-loop. In fact the 2 loop contribution $V_2$ leaves these
minima invariant. The very first check is the independence of the
gauge fixing procedure, and  this has been done$^9$. 

The formula 4.1 depends on one parameter q and describes only the
rim of the figure. The full expression for the inside of the base
triangle in fig. 4 is given in ref.9, including two loops.

In triangles neighbouring the one in fig. 3 one obtains the potential
by reflection. Hence the potential has a local minimum along this rim,
and the minimum configuration that gives us the surface tension is
precisely developing along this rim, from q=0(p=1)
to q=1(p=$exp{i2{\pi\over 3}}$). Strikingly, this is what is found in
MonteCarlo simulations. The result for the surface tension including
the two loop effects in eq.~4.1 is then~:
$$\alpha={4(N-1)\over 3\sqrt{3N}}{\pi}^2{T^3\over
g}(1-(15.27853...){g^2N\over {16{\pi}^2}})\eqno 4.2$$
Note the overall factor $1\over g$, not ${1\over g^2}$, for the lowest
order result. This follows immedia\-tely from eq.~4.1, and
that gradients are of order g. One just has to rescale the z variable
into $gz$.

\vskip 0.6cm
\leftline{\twelvebf 5. Conclusions}
\vglue 0.4cm

We have shown, that Z(N) bubbles are well alive,with the thickness of
the wall  $O({1/over {gT}})$. They are not an artifact of the
Euclidean path integral, since the surface tension can be defined
without mentioning the pathintegral. Including quarks lifts the
degeneracy of the vacua in fig. 4, see fig. 5. But there stays a
metastable state. This metastable state has received a lot of
attention, because of its strange thermodynamical properties$^{11}$. 

With quarks, there is no surface tension tension in the vein of
section 3. Gene\-rally speaking, the introduction of a field, that feels
the centergroup of the gauge group (quarks in QCD, electrons in QED)
will render the definition of states with a given flux as in section
3.1 impossible, since the field will become multivalued.

It has been noticed$^{12}$, that the 1-loop kinetic term contains a
pole at q=0. This means that for $q=O(g)$ perturbation theory for the
kinetic term  gets upset. But those values of q contribute only to
order $g^3$ to the surface tension eq.~4.2 as is readily seen, when
minimising eq.~4.1.

\vskip 0.6cm
\leftline{\twelvebf Acknowledgements}
\vglue 0.4cm

I'm indebted to Rob Pisarski, George Semenoff, Mischa Shifman, Andrei
Smilga, Mike Teper and Nathan Weiss for discussions. My thanks go to
Randy Kobes and
Gabor Kunststatter for organising a very inspiring meeting.

\vskip 0.6cm
\leftline{\twelvebf References}
\vglue 0.4cm

\itemitem{1.} K.~Kajantie, L.~Karkainen, K.~Rummukainen, Nucl. Phys.
B333, 100(1990).
\itemitem{  } S.~Huang, J.~Potvin, C.~Rebbi and S.~Sanielevici,
Phys. Rev. {\twelvebf D 42} (1990), 2864.
\itemitem{2.} T.~Bhattacharya, A.~Gocksch, C.P.~Korthals Altes and
R.D.Pisarski, Phys. Rev. Lett {\twelvebf 66}, 998 (1991). 
\itemitem{  } T.~Bhattacharya, A.~Gocksch, C.P.~Korthals Altes and
R.D.Pisarski, Nucl. Phys. {\twelvebf B383}, (1992), 497
\itemitem{3.} L.~MacLerran, B.~Svetitsky Phys. Rev. {\twelvebf D 24},
450.
\itemitem{  } J.~Kuti, J.~Polonyi, K.~Szlachanyi, Phys. Lett.
{\twelvebf 98B}, 199 (1981).
\itemitem{4.} G.~Muenster, Nucl. Phys. {\twelvebf B340} (1990), 559.
\itemitem{5.} G.'t Hooft, Nucl. Phys. {\twelvebf B153}, 141 (1979).
\itemitem{  } L.P.~Kadanof, H.~Ceva, Phys. Rev. {\twelvebf B3} (1971), 3918. 
\itemitem{6.} T.~Bhattacharya, A.~Gocksch, C.P.~Korthals Altes
and R.D.Pisarski, Nucl. Phys. {\twelvebf B (Proc. Suppl.)
20}, (1991), 305.
\itemitem{7.} G.~Muenster, P.~Weisz, Phys. Lett. {\twelvebf 96B},
(1980), 119.
\itemitem{8.} K.~Binder, Z. Phys. {\twelvebf B43} (1981), 119.
\itemitem{9.} C.P.~Korthals Altes, Marseille preprint, {\twelvebf
CPT-93}/P.2928.
\itemitem{10.} D.~Gross, R.D.~Pisarski, L.G.~Yaffe, Rev. Mod. Phys.
{\twelvebf 53}, 43 (1981).
\itemitem{   } N.~Weiss, Phys. Rev. {\twelvebf D~24}, 475 (1981).
\itemitem{11.} V.~Dixit, M.~Ogilvie, Phys. Lett. {\twelvebf B269},
353 (1991)
\itemitem{   } V.M.~Belyaev, I.I.~Kogan, G.W.~Semenof, N.~Weiss, Phys.
Lett. {\twelvebf B277}, 331 (1992)
\itemitem{12.} A.V.~Smilga, Bern University preprint {\twelvebf
BUTP-93}/3.

\vfill\eject

%\input epsf
%\def\epsfsize#1#2{0.8#1}

%\null

%\hskip 3,6truecm\vbox{\vskip 7,5truecm % (Taille de la figure en cm)
%\special{illustration fig1.ps scaled 800}}
%\epsfbox{/home/cptsu1/nicolis/Net/Test/fig1.ps}

\vskip 1truecm

Fig.1
\medskip
Classical Potential in $\lambda\varphi^4$ with spontaneous
symmetry breaking.

%\vfill\eject

%\null

%\hskip 4,3truecm\vbox{\vskip 7,5truecm 
%\special{illustration fig2.ps scaled 800}}
%\epsfbox{/home/cptsu1/nicolis/Net/Test/fig2.ps}

\vskip 1truecm

Fig. 2
\medskip
{\parindent=1truecm\narrower
\noindent System with fixed boundary conditions $\varphi=\pm v$. The
correlation function has endpoints $x$ and $y$.\par}

%\vfill\eject

%\null

%\hskip 0,6truecm\vbox{\vskip 7,5truecm 
%\special{illustration fig3.ps scaled 800}}
%\epsfbox{/home/cptsu1/nicolis/Net/Test/fig3.ps}

%\vskip 1truecm

Fig. 3
\medskip
{\parindent=1truecm\narrower
\noindent Deformation of the dislocation line (dotted) in the
presence of the correlation (the crosses). Horizontal bars are
twisted links.\par}

%\vfill\eject

%\null

%\hskip 2,3truecm\vbox{\vskip 4,5truecm 
%\special{illustration fig4.ps scaled 800}}
%\epsfbox{/home/cptsu1/nicolis/Net/Test/fig4.ps}

\vskip 1truecm

Fig. 4
\medskip
One loop pure glue potential for $N=3$.

%\vfill\eject

%\null

%\hskip 2,3truecm\vbox{\vskip 4,5truecm 
%\special{illustration fig5.ps scaled 800}}
%\epsfbox{/home/cptsu1/nicolis/Net/Test/fig5.ps}

%\vskip 1truecm

Fig. 5
\medskip
As in fig. 4, but with two quark flavours.

\vfill\eject
\bye
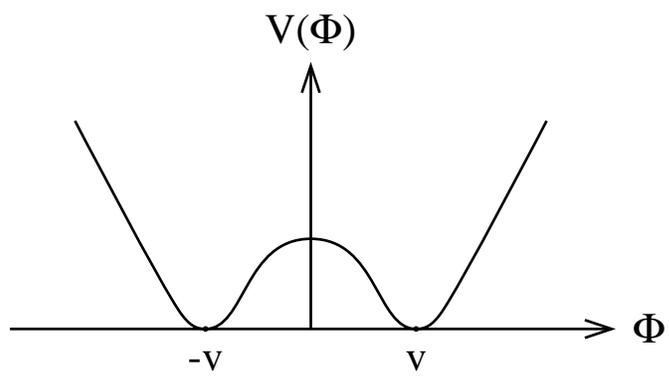
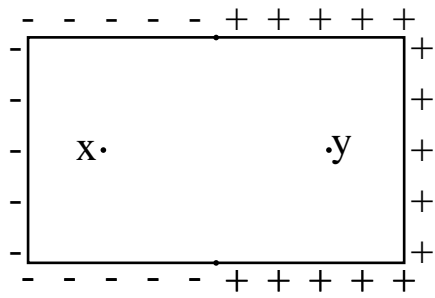
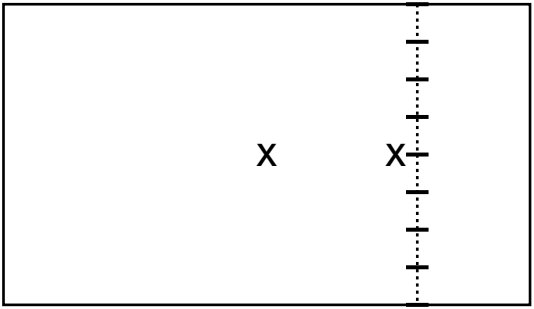
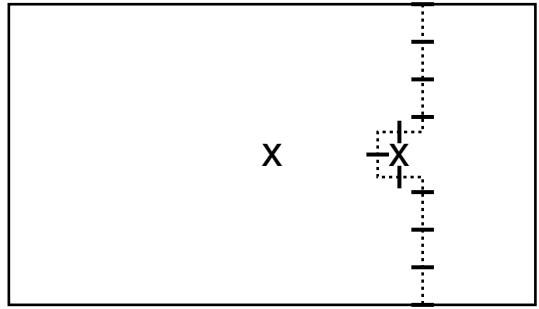
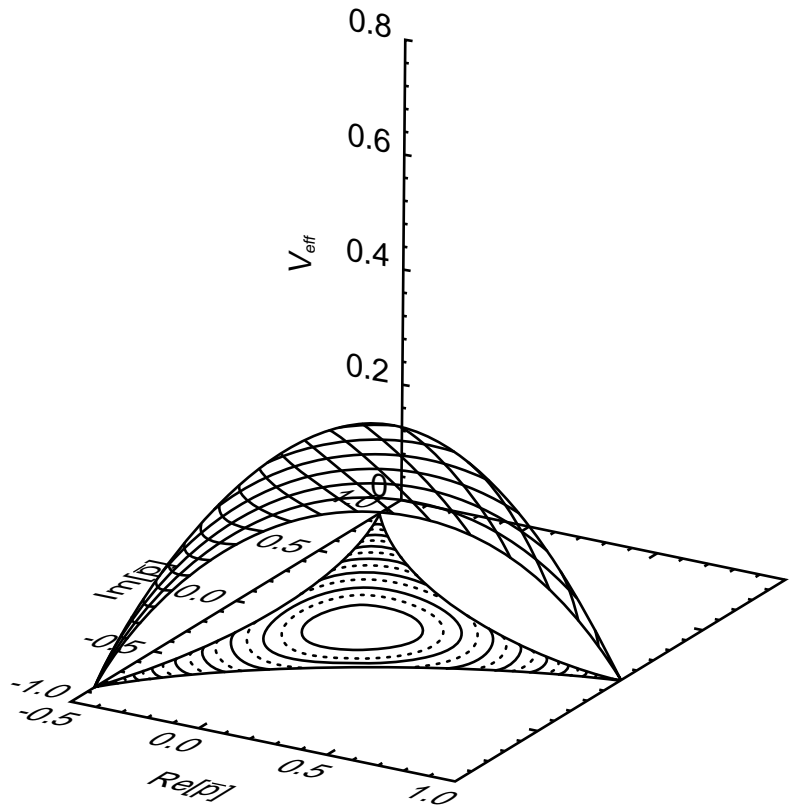
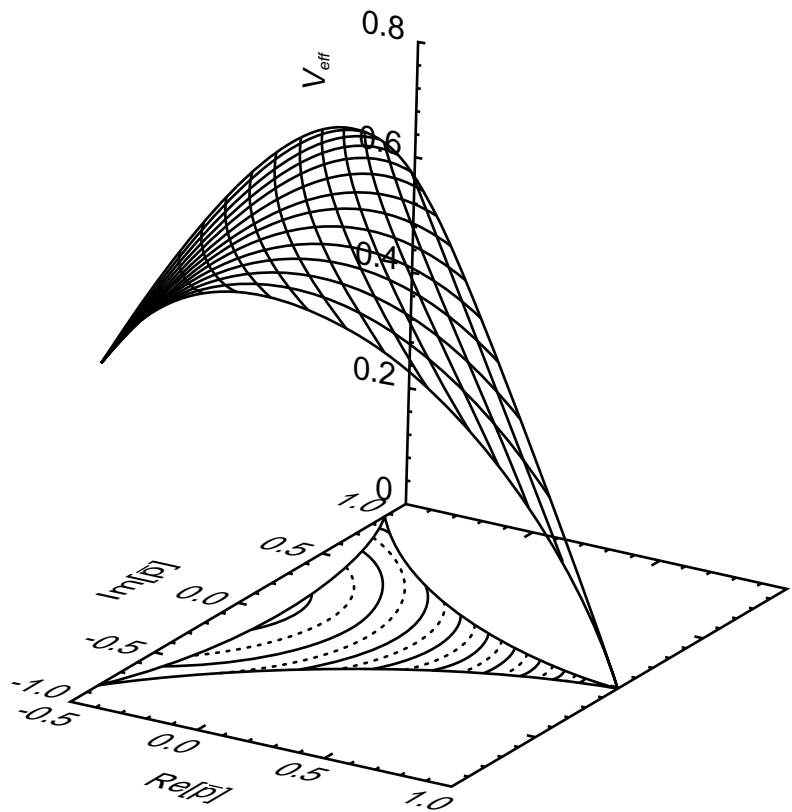
\end